# Observation of Low Energy Raman Modes in Twisted Bilayer Graphene

*Rui He,*[†,⊥,*] *Ting-Fung Chung,*[‡,⊤,⊥] *Conor Delaney,*[†] *Courtney Keiser,*[†] *Luis A. Jauregui,*[⊤,∥] *Paul M. Shand,*[†] *C. C. Chancey,*[†] *Yanan Wang,*[§] *Jiming Bao,*[§] *Yong P. Chen*[‡,⊤,∥,*]

[†] Department of Physics, University of Northern Iowa, Cedar Falls, IA 50614, USA

[‡] Department of Physics, Purdue University, West Lafayette, IN 47907, USA

[⊤] Birck Nanotechnology Center, Purdue University, West Lafayette, IN 47907, USA

[∥] School of Electrical and Computer Engineering, Purdue University, West Lafayette, IN 47907, USA

[§] Department of Electrical and Computer Engineering, University of Houston, Houston, TX 77204, USA

[⊥] These authors equally contributed to this work.

[*] Address correspondence to rui.he@uni.edu;   yongchen@purdue.edu

ABSTRACT: Two new Raman modes below 100 cm$^{-1}$ are observed in twisted bilayer graphene grown by chemical vapor deposition. The two modes are observed in a small range of twisting angle at which the intensity of the G Raman peak is strongly enhanced, indicating that these low energy modes and the G Raman mode share the same resonance enhancement mechanism, as a function of twisting angle. The ~94 cm$^{-1}$ mode (measured with a 532 nm laser excitation) is assigned to the fundamental layer breathing vibration (ZO' mode) mediated by the twisted bilayer graphene lattice, which lacks long-range translational symmetry. The dependence of this mode's frequency and linewidth on the rotational angle can be explained by the double resonance Raman process which is different from the previously-identified Raman processes activated by twisted bilayer graphene superlattice. The dependence also reveals the strong impact of electronic-band overlaps of the two graphene layers. Another new mode at ~52 cm$^{-1}$, not observed previously in the bilayer graphene system, is tentatively attributed to a torsion mode in which the bottom and top graphene layers rotate out-of-phase in the plane.

TEXT:

In bilayer graphene (BLG) exfoliated from highly oriented pyrolytic graphite (HOPG) crystals, the top and bottom graphene layers usually form Bernal (A-B) stacking in which the carbon atoms of the B sublattice of the second layer sit on top of the A sublattice carbon atoms of the first graphene layer. Two parallel parabolic conduction bands situated above another two parallel parabolic valence bands, with zero bandgap, characterizes the low energy (near the charge neutral point) electronic band structure of Bernal-stacked BLG.[1-3] However, in twisted BLG (tBLG) in which the second graphene layer is rotated with respect to the first layer, the low energy electronic band structure can be represented by two Dirac cones separated by a wavevector that depends on the rotational angle.[4, 5] While linear dispersion is maintained in tBLG system, van Hove singularities in the density of states (DOS) are generated due to the coupling of the two layers.[4] In addition, tBLG exhibits interesting optical features with optical absorption bands in the visible range.[6,7] Hence, probing the fundamental properties of tBLG is of interest and importance.

Phonons play an important role in electron transport in BLG through electron-phonon interactions.[8] Low energy phonons, *e.g.* the layer breathing mode in which the two graphene layers vibrate out-of-phase perpendicular to their planes, facilitate interlayer current conduction in tBLG.[9] Raman spectroscopy is a noninvasive, direct, and sensitive probe of phonons in graphene layers. It has been shown that the intensity of the G Raman peak and the position, linewidth, and intensity of the 2D Raman band undergo characteristic changes as a function of twisting angle.[4,5] In addition, new Raman lines, *e.g.* R and R' peaks close to the G band, other lines around the D peak, and out-of-plane acoustic (ZA) and layer breathing (ZO') modes (between 120 and 200 cm$^{-1}$) are observed in tBLG due to Raman processes in which finite wavevector phonon scatterings are activated by involving angle dependent superlattice wavevectors.[10-12]

In this paper, we report the observation of new low energy vibrational modes below 100 cm$^{-1}$ in Raman scattering from tBLG. For a given laser excitation energy, these modes are only observed in the vicinity of a specific twisting angle at which the G Raman band is strongly enhanced. This observation indicates that the resonance enhancement mechanism of the low energy modes is similar to that of the G band enhancement.[4,5] In addition to the resonance enhancement of the intensities of the observed low energy Raman modes, the intensity of the background on which the low energy Raman peaks are superimposed is also enhanced near the critical twisting angle. This broad and enhanced background at low energy could be related to electronic excitations in tBLG. The Raman peak at ~94 cm$^{-1}$ (measured with laser wavelength 532 nm), referred to as (ZO')$_L$ mode in this paper, is assigned to the fundamental layer breathing mode (arising from the out-of-plane relative motions of the two graphene layers). This mode can be explained by the double resonance mechanism and may be activated by the tBLG lattice which lacks long-range translational symmetry. The dependence of the frequency and linewidth of this mode on the twisting angle reveals the degree of overlap of the two Dirac cones that belong to the two layers in tBLG. This fundamental ZO' mode is also observed at a higher frequency above 100 cm$^{-1}$ (here referred to as (ZO')$_H$) and is understood to be mediated by the tBLG superlattice with wavevector *q*.[13] The coexistence of two fundamental ZO' Raman lines originating from phonons in different parts of the Brillouin zone (with different phonon wavevectors) suggests that although the tBLG system lacks long range translational symmetry, superlattice periodicity can still be defined. In contrast, in Bernal-stacked BLG, the fundamental layer breathing (ZO') mode is known to be spectroscopically inactive (silent) and has never been observed previously (only its overtone 2ZO' was observed).[14, 15] Our observation indicates that tBLG differs dramatically from Bernal-stacked BLG, and highlights that tBLG is a very interesting system which allows us to probe a broad range of phonon dispersion in the interior of the Brillouin zone. Another mode at ~52 cm$^{-1}$ is tentatively assigned to a torsion mode that also appears to be activated in the tBLG system but not in the Bernal-stacked BLG.

Our graphene layers were grown by chemical vapor deposition (CVD) on Cu foils at ambient pressure[16, 17] (APCVD) and transferred onto a highly p-doped Si substrate (with ~300 nm SiO$_2$) for all subsequent measurements. Details of the sample growth are shown in the Supporting Information. Figure 1(a) shows an optical image of CVD graphene on Si/SiO$_2$ substrate. Single layer and bilayer graphene areas can be identified from the color contrast. It is seen that ~70-80% of the substrate is covered by graphene (polycrystalline consisting of single crystal graphene domains). The first layer domains have a typical size of ~20 μm and largely connect with neighboring domains. The second layer domains are often located near the centers of the first layer domains and have a typical lateral size of a few microns. The graphene domains grown by our CVD method are mostly hexagonal in shape, with edges parallel to zigzag directions of the graphene lattice[17] (thus facilitating the determination of the lattice orientation using these edges; see below). The number of layers is confirmed by both Raman measurements as well as by measurements of the thickness of bilayer domains by atomic force microscopy (AFM) in a tapping mode (see a representative AFM image in the Supporting Information).

Figure 1(b) illustrates three representative Raman spectra measured from tBLG domains with different twisting angles. From the G and 2D Raman band features,[4,5] we can estimate the twisting angles. The insets display the corresponding optical images of each BLG domain. Contours of the single-layer (dashed lines) and bi-layer (solid lines) areas have been included as guides to the eye. The twisting (misorientation) angles can also be determined by measuring the angles formed between neighboring edges of the first and the second layer domains.[18] Both methods provide consistent estimation of twisting angles (within ~2°). Figure 1(c) summarizes the distribution of twisting angles determined by the Raman method. A similar histogram of BLG twisting angles is also obtained by measuring the relative orientations between the edges of top and bottom graphene layers seen in optical images (see the Supporting Information). The twisting angle distribution suggests that our APCVD growth favors tBLG with a large twisting angle (20°–30°). This is very different from the bilayer domains grown by low-pressure (LP) CVD that are largely in Bernal-stacked configuration (no-twisting).[19,20] This finding reveals that the distribution of rotational angles in tBLG depends on the growth conditions, which influence the growth kinetics of CVD graphene.

Figure 2(a) shows five representative Raman spectra measured down to low Raman shifts (<100 cm$^{-1}$) from our probes of five different bilayer domains. A spectrum from single layer graphene (SLG) is included for comparison. From the positions, linewidths and intensities of the R, G and 2D Raman bands, we can approximately determine the twisting angles in these bilayer domains (labeled in Fig. 2(a)). For 532 nm laser excitation, the G peak intensity reaches maximum and concomitantly the 2D band exhibits the greatest blueshift from that of the SLG when the twisting angle $\theta$ is near the critical twisting angle $\theta_c = 12°$ (see the red spectrum in Fig. 2(a)).[5] Qualitatively similar phenomena are seen for the 633 nm laser excitation (see the Supporting Information). The critical angle $\theta_c$ varies for different laser excitation energies.[4] This is shown by different intensities in Raman mapping over the same tBLG island by using different laser excitation energies[21] (also see the Supporting Information). All Raman data shown in the main text are measured with a 532 nm laser excitation unless otherwise stated. The D peak intensity is very low or negligible in most bilayer domains that we studied. This implies that the quality of the graphene layers is high. In addition to the G and 2D peaks, several low energy Raman modes and a strong background on which the low energy Raman modes superimpose are observed when the twisting angle is in the vicinity of $\theta_c$. A representative zoomed-in low-energy spectrum is shown in the left inset of Fig. 2(a). The inset on the right-hand side shows the same spectrum after subtracting the background (highlighted by the dashed line in the left inset). Four Raman modes (as shown by 4 Lorentzian peaks after the spectrum is decomposed) are observed within the range of 30-200 cm$^{-1}$. These low energy Raman features, particularly the two newly-observed modes below 100 cm$^{-1}$, are the main focus of this paper.

Dispersion of low-energy phonons in tBLG has not been well-explored thus far. Theoretical calculation of dispersion curves of low-energy phonons of the tBLG system is very challenging since the unit cell size required is very large. Only limited experimental work of low energy phonons (in the range 100-200 cm$^{-1}$) and dispersions has been reported in the literature.[13] These phonons of above 100 cm$^{-1}$ are described by Raman scattering processes mediated by the superlattice wavevector *q* which depends only on the twisting angle $\theta$.[10, 13] The frequency-wavevector relation for these phonons overlaps with the SLG and the ZO' (layer breathing mode as schematically shown in Fig. 4(b)) phonon dispersion curves of Bernal-stacked BLG.[13] Figure 5 shows low-frequency (below 200 cm$^{-1}$) phonon dispersion.

Figures 2(b) and (c) show original and background-subtracted low-energy Raman spectra from several different bilayer domains. From the R, G and 2D Raman characteristics (spectra are shown in the Supporting Information), we determined that the twisting angles $\theta$ of these bilayer domains range from ~11° (< $\theta_c$) to ~14° (> $\theta_c$). Two modes highlighted by asterisks and squares are observed between 130-180 cm$^{-1}$ (see Fig. 2(c)). The frequencies of these two modes blueshift monotonically with the increase of the twisting angle $\theta$ and agree well with those reported in Ref. [13]. Hence, these modes highlighted by asterisks and squares are attributed to fundamental layer breathing (ZO') and out-of-plane acoustic (ZA) modes, respectively.[13] They are activated by the superlattice (Moire pattern) formed in tBLG.[10] The momentum conservation condition is satisfied by the participation of superlattice wavevector *q* (see Fig. 4(a)). The magnitudes of the scattered phonons thus equal that of the superlattice wavevector *q*, which depends on the twisting angle $\theta$ and is about 0.65/0.54 Å$^{-1}$ for a twisting angle near $\theta_c$ (≈12°/10°) for 532/633 nm laser excitation. The points circled in Fig. 5 show the frequencies and wavevectors of these ZO' and ZA phonons activated by the tBLG superlattice as observed in our work.

Figures 2(b) and (c) show that in addition to the modes between 130 and 180 cm$^{-1}$ two even lower frequency modes are observed. The lowest observed frequency mode (labeled "*X*") occurs at ~52 cm$^{-1}$. The next higher frequency mode is observed at ~94 cm$^{-1}$, which is close to the frequency of the ZO' mode calculated for and inferred from the observed overtone (2ZO') in Bernal-stacked bilayer graphene.[15, 22, 23] Hence, we assign this mode to another fundamental layer breathing mode ZO' whose phonon wavevector *q'* is different from the superlattice wavevector *q*. Because this ZO' frequency is lower than that induced by the superlattice, we name this ZO' at 94 cm$^{-1}$ as (ZO')$_L$ and the ZO' at higher energy of above 100 cm$^{-1}$ as (ZO')$_H$.

It has been shown by Kim *et. al.* that the integrated intensity of the 2D Raman peak of tBLG increases monotonically with the twisting angle *θ* in the vicinity of $θ_c$.[4] Their studies were conducted on suspended tBLG by consecutively transferring two SLG on a carbon TEM grid where the effects due to the substrates (*e.g.* doping) are minimized. The same relation between 2D Raman integrated intensity and twisting angle is also observed in our samples (see Figs. S2-S4 in the Supporting Information). This suggests it is reasonable to use the integrated intensity of the 2D peak to characterize the twisting angle of our samples. Therefore, we summarize our measurements of the (ZO')$_L$ mode by plotting the position, bandwidth, and integrated intensities of this mode as a function of normalized 2D Raman intensity ($I_{2D}$, normalized to the intensity of SLG, *i.e.* take the ratio of 2D integrated intensity of BLG to that of the SLG on the sample). Because the G-peak position is sensitive to doping,[8, 24] we also plot the G-peak frequency as a function of normalized $I_{2D}$. The G-peak frequency vs. normalized $I_{2D}$ plot (see Supporting Information) is very similar to that obtained in suspended tBLG samples reported by Kim *et. al.*[4] This further confirms that doping by the substrate is not a major concern in our studies of the low energy modes as a function of twisting angle. By comparing the Raman spectra of our SLG domains on the same substrate with previous Raman studies of strained graphene,[25, 26] we confirm that strain is not substantial in our samples.

Figures 3(a) and (b) display the evolution of frequency and full-width-at-half-maximum (FWHM) of the $(ZO')_L$ mode as a function of normalized $I_{2D}$. The range from 1.1 to 2.2 of the normalized $I_{2D}$ corresponds to a range of twisting angle $\theta$ from ~10° to ~15°. It is seen that the frequency of this $(ZO')_L$ mode ($\omega_{(ZO')_L}$) increases with increasing normalized $I_{2D}$ when the normalized $I_{2D}$ is below 1.5 (or when $\theta$ <12°).[4] The frequency $\omega_{(ZO')_L}$ becomes almost constant after the normalized $I_{2D}$ is greater than 1.5 (or when $\theta$ > 12°). Figure 3(b) shows that the FWHM of the $(ZO')_L$ mode decreases with increasing normalized $I_{2D}$ when it is below 1.5 ($\theta$ < 12°) and that it also becomes nearly constant when the normalized $I_{2D}$ is greater than 1.5 ($\theta$ > 12°). These results indicate that the dramatic transitions in the frequency and FWHM of the $(ZO')_L$ mode occur when the twisting angle $\theta$ is near $\theta_c$. In order to confirm the characteristics of the $(ZO')_L$ mode, we also plot (see Supporting Information) the position and width of this peak as a function of the R mode frequency which decreases monotonically with $\theta$.[13] Similar trends as those plotted as a function of normalized $I_{2D}$ are observed.

The coexistence of two ZO' phonons with different wavevectors is a novel phenomenon. Because the $(ZO')_H$ phonon's wavevector is defined by the tBLG superlattice, the fundamental $(ZO')_L$ mode that we observe for the first time in Raman scattering from tBLG must be activated by a different wavevector that satisfies the momentum conservation requirement. We propose that the $(ZO')_L$ mode at ~94 cm$^{-1}$ is facilitated by the tBLG crystal lattice which lacks translational symmetry. The Raman process for this $(ZO')_L$ phonon is an intravalley scattering process[15, 23] and involves four steps (see Figs. 4(c)-(e)): (i) The incident photon creates an electron-hole pair; (ii)

The electron/hole is scattered by a tBLG crystal lattice (shown by the dashed arrow which provides a momentum $q'$); (iii) The electron/hole is scattered by a phonon with wavevector $k_{(ZO')_L}$; (iv) Electron-hole recombination. In $k$-space, the Dirac cones from top and bottom graphene layers (located at $K_a$ and $K_b$) are separated from each other by a distance that depends on the twisting angle $\theta$ (see Fig. 4(a)).[4, 10] The larger the $\theta$, the more separated the $K_a$ and $K_b$ points, and thus more distant the two Dirac cones are. When $\theta = \theta_c$ (see Fig. 4(d)), the incident photon energy $\hbar\omega_{in}$ equals the energy difference between the conduction and valence van Hove singularities.[4, 5] Because tBLG maintains linear electronic dispersion of SLG, i.e. $E = \hbar v_F k$, where the Fermi velocity $v_F \approx 1\times10^6$ m/s and $E$ = 2.33 eV for a 532 nm photon, we can estimate the magnitude of the phonon wavevector $k_{(ZO')_L} = q' = 0.36$ Å$^{-1}$ under 532 nm laser excitation. In this case, the magnitude of $k_{(ZO')_L}$ or $q'$ is roughly half of the superlattice wavevector $q$ and equals the distance between $K_a$ and $K_b$ (as shown in Figs. 4(a) and (d)). The observed phonon frequency (94 cm$^{-1}$) and the phonon wavevector $k_{(ZO')_L}$ are in excellent agreement with the ZO' phonon dispersion in Bernal-stacked BLG (see the two uncircled points in Fig. 5 for 532 nm and 633 nm laser excitations),[23, 27-29] which confirms our assignment of this mode to the layer breathing mode $(ZO')_L$. It is worthwhile noting that momentum conservation in the Raman process is achieved through an intermediate step shown by the dashed arrow in Fig. 4(d). It is unlikely that this step is due to defect-induced scattering since the Raman intensity of the D band is reasonably weak in our samples (see the full Raman spectra in Fig. 2(a) and in the Supporting Information). Therefore, this intermediate step is likely due to scattering of electrons by the tBLG crystal lattice, which lacks long range periodicity (lattice translational symmetry).[10, 19] This activation process is confirmed by the absence of the fundamental $(ZO')_L$ mode Raman scattering from Bernal-stacked

BLG that has long-range lattice translational symmetries.[15, 23, 30] In Bernal-stacked BLG in which scattering due to tBLG crystal lattice is absent, the second-order (2ZO') mode in which momentum conservation is satisfied by involving two ZO' phonons with opposite wavevectors, displays strong Raman intensity.[15] In addition, this (ZO')$_L$ phonon in tBLG softens under 633 nm laser excitation (see Supporting Information), which further confirms that this phonon has a nonzero wavevector and that the scattering process can be explained by the double resonance mechanism.

When $\theta < \theta_c$, the two Dirac cones have substantial overlap. The Raman process accesses the overlapped area (involving both Dirac cones at locations $K_a$ and $K_b$) and allows phonons with wavevector $k_{(ZO')_L}$ less than 0.36 Å$^{-1}$ to contribute to the Raman spectra (see Fig. 4(c)). In this framework, the FWHM of this (ZO')$_L$ mode will increase as $\theta$ decreases (as the area of overlapped Dirac cones increases), which is supported by our observation shown in Fig. 3(b). In addition, the frequency of this (ZO')$_L$ mode should reduce slightly as the twisting angle $\theta$ decreases away from $\theta_c$ as the overlapped Dirac cones enable (ZO')$_L$ phonon with wavevector $k_{(ZO')_L}$ less than 0.36 Å$^{-1}$ to contribute to the Raman spectra, which would lower the phonon energy (based on the ZO' phonon dispersion predicted in BLG[23, 27-29]). Our observation of (ZO')$_L$ phonon softening and broadening for $\theta < \theta_c$ (see Figs. 3(a) and (b)) confirms this interpretation and shows that the phonon wavectors $k_{(ZO')_L}$ (or $q'$) is not uniquely defined by the twisting angle $\theta$. For $\theta < \theta_c$, $k_{(ZO')_L}$ and $q'$ can have multiple values depending on the degree of overlap of the two Dirac cones (see Fig. 4(c)). This is very different from the superlattice wavevector $q$ which is solely determined by the twisting angle $\theta$.

For $\theta > \theta_c$, the Raman process does not access overlapped Dirac cones (see Fig. 4(e)). Even if the area of Dirac cone overlap decreases as $\theta$ increases, it should not affect the $(ZO')_L$ phonon wavevector involved in the Raman process. This agrees well with our observation that the frequency and FWHM of the $(ZO')_L$ mode remain unchanged when the twisting angle $\theta$ is greater than $\theta_c$, as shown in Figs. 3(a) and (b). These results indicate that the $(ZO')_L$ phonon is not very sensitive to the twisting of the two graphene layers for $\theta \geq \theta_c$. This feature of the $(ZO')_L$ mode is very different from that of the $(ZO')_H$ and other Raman modes, *e. g.* R and R', whose frequencies vary monotonically with the twisting angle $\theta$ and are sensitive to $\theta$ for both $\theta > \theta_c$ and $\theta < \theta_c$.[13] This difference indicates that the $(ZO')_L$ mode does have a different Raman scattering mechanism from those phonons which are activated by the tBLG superlattice. The double resonance mechanism that we proposed above qualitatively explains the characteristics of the $(ZO')_L$ mode as a function of $\theta$.

The tBLG system allows us to probe the dispersion of the layer breathing mode ZO' in a broad range off the Brilluoin zone center (see Fig. 5). Our measurements show that the dispersion of the ZO' mode in tBLG is similar to that of Bernal-stacked BLG,[15] which indicates that the interlayer *out-of-plane* vibrations of tBLG are comparable to those of Bernal-stacked BLG for $\theta \sim \theta_c$, consistent with the calculation shown in Ref. [31]. The emergence of this *fundamental* layer breathing (ZO') vibration in the *twisted* BLG system (known to be silent in Bernal-stacked BLG and graphite[14]) implies that the crystal symmetry that makes the ZO' silent in Bernal-stacked BLG is lifted in tBLG.

Figure 3(c) displays the change of integrated intensity of the $(ZO')_L$ mode as a function of normalized $I_{2D}$ (and thus as a function of $\theta$). It is seen that the $(ZO')_L$ mode intensity is strongly enhanced when the normalized $I_{2D}$ is ~1.6, consistent with the critical value of 1.5 found in the changes in the frequency (Fig. 3(a)) and FWHM (Fig. 3(b)) of this mode. The normalized $I_{2D}$ of ~1.6 corresponds to a twisting angle $\theta \approx 12.5°$, which agrees well with the value of $\theta_c$.[5] This observation indicates that the Raman intensity of the $(ZO')_L$ phonon displays large resonance enhancement at $\theta_c$, where the intensity of the G Raman peak is also enhanced (see Fig. 3(e) and Refs. 4 & 5). Our observation suggests that the $(ZO')_L$ and the G modes share the same resonance enhancement mechanism arising from van Hove singularities in the DOS in the tBLG system.[4, 5] The ratio of integrated intensities of the $(ZO')_L$ and G peaks reaches a maximum value of ~7% at resonance. It is very impressive that when $\theta$ overlaps $\theta_c$ the intensity of the $(ZO')_L$ peak is comparable to that of the G band (off resonance), and that it is much stronger than the intensities of ZA, $(ZO')_H$, R and R' peaks, as shown in Fig. 2. This difference in resonance enhancement between the $(ZO')_L$ and other modes further confirms that the Raman scattering mechanisms of these modes are different. The $(ZO')_L$ mode which involves optical transitions between the conduction and valence van Hove singularities (see Fig. 4(d)) are strongly enhanced, whereas the ZA, $(ZO')_H$, R and R' modes due to superlattice scattering are not subject to the same enhancement. When $\theta$ differs from $\theta_c$ significantly ($\theta < 10°$ or $> 15°$), this $(ZO')_L$ mode is not observed, likely due to its intensity becoming too weak to be detected.

Another striking feature of the $\theta$-dependent Raman spectra from tBLG is the large enhancement of the background intensity of low energy Raman modes at the critical twisting angle $\theta_c$. The background envelope on which the low-energy Raman lines superimpose (as highlighted by dashed lines in Fig. 2(b)) is steepest for $\theta \sim \theta_c$. We define the height intensity of this background envelope at 70 cm$^{-1}$ as the low-energy background intensity (as illustrated by the vertical arrow for the spectrum with $\theta \sim 12°$ in Fig. 2(a)). Figure 3(d) shows the change of this low-energy background intensity as a function of the normalized $I_{2D}$ (and thus as a function of $\theta$).

It is seen that this low-energy background intensity also reaches maximum at $\theta_c$ (when the normalized $I_{2D}$ is ~1.6), similar to the resonance enhancements of the $(ZO')_L$ and G Raman peaks. We speculate that this broad low-energy background envelope which is strongly enhanced at the critical angle $\theta_c$ could be related to electronic excitations such as plasmons in tBLG.[32, 33] Further investigations are needed to understand the origin of this enhanced low energy background.

Figure 2 shows that in addition to the two ZO' modes and ZA mode, an even lower energy mode around 52 cm$^{-1}$ (*X* mode) is observed for $\theta \sim \theta_c$. This "*X*" mode appears to exhibit similar resonance enhancement as the $(ZO')_L$ and G modes do near the critical twisting angle $\theta_c$, and may be too weak to be seen when $\theta$ is off $\theta_c$. Like the two ZO' modes, this *X* mode is also only observed in tBLG, and is not observed in Bernal-stacked BLG.[30] On the other hand, previous experiments on Bernal-stacked BLG observed a shear mode (*i.e.*, the *C* mode, occurring at ~31 cm$^{-1}$) that originates from the relative in-plane sliding of the two graphene layers.[30] An interlayer coupling strength of ~ $12.8 \times 10^{18}$ Nm$^{-3}$ is estimated from the position of the *C* mode in Bernal-stacked graphene layers.[30] In our measurements of tBLG, the rising background below 50 cm$^{-1}$ (see Figs. 2(a) and (b)) makes it very challenging to resolve and investigate the *C* mode (usually very weak in bilayer graphene) in our samples. Hence, it is hard for us to estimate the interlayer coupling constant from the *C* mode of tBLG and compare it to that of Bernal-stacked graphene or graphite. The frequency of this *X* mode is much higher than that of the ZA mode with the phonon wavevector of ~0.36 Å$^{-1}$ (see Fig. 5). Therefore, the *X* mode cannot be assigned to a ZA phonon due to double resonance scattering, unlike the $(ZO')_L$ mode. The exact nature of the *X* mode that we observe in tBLG is not yet clear at this time. We believe that the *X* mode is a different mode from the *C* mode since the position, linewidth, and lineshape of the two modes are very different. The *C* mode in Bernal-stacked graphene has a narrow (a few cm$^{-1}$ in width) asymmetric Fano lineshape that results from quantum interference between a discrete phonon state and a relatively broad continuum of electronic or multiphonon transitions.[30] In contrast, the *X* and (ZO')$_L$ modes in tBLG have larger FWHM (10-15 cm$^{-1}$) and Lorentzian lineshapes (see Fig. 2), which are similar to those of the G Raman peak. This observation suggests that the *X* and (ZO')$_L$ peaks are mainly influenced by the electron-phonon coupling, similar to what occurs for the G band.[8, 30]

We suggest that one possible assignment for the *X* mode may be a torsional motion in which the top and bottom graphene layers execute out-of-phase rotations as schematically shown in Fig. 4(b). There will be multiple approximate symmetries for various twisting angles $\theta$ ($0° \leq \theta \leq 30°$), and the potential energy landscape should be characterized by small energy barriers between adjacent configurations, promoting low-energy torsional motions about a particular $\theta$.[34] Analogous low-energy motions are seen in C$_{60}$, where electron-phonon interactions give rise to low-energy torsional motions about distortion configurations.[34] In the tBLG system, the observed resonance enhancement of Raman scattering intensity from this torsion mode near $\theta_c$ could result from electron-phonon coupling.

Such torsional modes in layered materials have received very little investigation so far. Studies of torsional modes in spherical materials such as nanoparticles have shown that if the shape of the particles is asymmetric due to deformation, torsional modes can be observed in low-frequency Raman scattering.[35, 36] Based on this scenario, we speculate that the presence of a torsional mode in tBLG could be related to the lack of long-range translational symmetry in the system. The apparent absence of torsional modes in Bernal-stacked BLG may follow simply from the absence of rotations in the translational symmetry group: the energy cost of a small-angle rotation is greater than shear mode translations.[34]

Unlike the ZO′ mode which is an out-of-plane vibrational mode, the relative in-plane motion, such as the $C$ mode and the torsional mode, of tBLG is expected to exhibit significant changes in comparison to that in Bernal-stacked graphene layers since tBLG has very little potential energy barrier to relative interlayer motion, as evidenced by superlubricity of rotated graphite[37] and multilayer graphene system.[19] The interlayer coupling in tBLG relevant for the $X$ (torsion) mode could be linked to a restoring force that tends to rotate the twisted graphene layers to their natural stacking structures[19] and/or Moire periodic potentials formed in two rotated honeycomb lattices.[25, 38-40] The presence of the $X$ (torsion) mode and the fundamental layer breathing mode suggests that the two layers in tBLG do couple to each other, consistent with the existence of interlayer transport in tBLG.[9, 41, 42]

In summary, two Raman modes below 100 cm$^{-1}$ previously unobserved in Bernal-stacked BLG are seen in tBLG when the twisting angle is close to the critical angle at which the intensity of the G Raman peak is enhanced. The mode observed at ~94 cm$^{-1}$ (measured with a 532 nm laser) is assigned to the fundamental layer breathing mode $(ZO')_L$. The intensities of this $(ZO')_L$ mode and the background envelope on which the low energy Raman lines superimpose display large resonance enhancements near the critical angle, concomitant with the large enhancement of the G Raman peak. The changes in position and linewidth of this mode as a function of twisting angle can be explained by the double resonance mechanism. It reveals the influence of angle-dependent electronic band overlaps on the Raman spectra. Another higher frequency mode $(ZO')_H$ induced by superlattice modulation is also observed in Raman spectra simultaneously. The other low-energy Raman mode observed at ~52 cm$^{-1}$ is tentatively attributed to the torsion mode in which the two graphene layers rotate in the plane with respect to each other. It is worthwhile noting that although our proposed double resonance mechanism explains the Raman features of the $(ZO')_L$ mode, there are two questions that remain open. One is the absence of the overtone $2(ZO')_L$ mode (expected to occur at ~188 cm$^{-1}$) in the Raman spectra from tBLG (see Figs. 2(b) and (c)). The second is why only the $(ZO')_L$ line is observed yet no phonons from other branches are observed in the double resonance Raman process. We speculate that this could be linked to the nonzero electron-phonon matrix element of the ZO' vibration.[28] Further studies are required to understand the mechanism causing the absence of $2(ZO')_L$ and other phonons within the double resonance framework in tBLG. A recent theoretical investigation has shown that phonon behaviors in tBLG are rather complex.[43] Our studies demonstrate that twisted bilayer graphene is a new system which exhibits fundamental properties that are distinct from those of Bernal-stacked bilayer graphene.

Supporting Information: The material contains details about sample preparation, Raman measurement conditions, AFM characterization of BLG domains, characterization of twisting angles, additional Raman characterizations of tBLG domains including the characterization of spectra excited by the 633 nm laser light, and Raman maps of the low-energy mode from a tBLG domain.

ACKNOWLEDGMENTS: R. H. acknowledges support by a Provost's Pre-Tenure Summer Fellowship Award from the University of Northern Iowa. R. H. also acknowledges the donors of The American Chemical Society Petroleum Research Fund (Grant No. 53401-UNI10) for partial support of this research. C. D. acknowledges the support of a SOAR award made by UNI College of Humanities, Arts & Sciences. P. M. S. acknowledges support by NSF Grant No. DMR 1206530. J. M. B. acknowledges the support from National Science Foundation (ECCS-1240510 and DMR-0907336) and the Robert A Welch Foundation (E-1728). Y. P. C. thanks A. C. Ferrari and M.-Y. Chou for helpful discussion. Y. P. C. acknowledges partial supports from NSF, DTRA and C.F. Day & Associates for the graphene synthesis and characterizations.

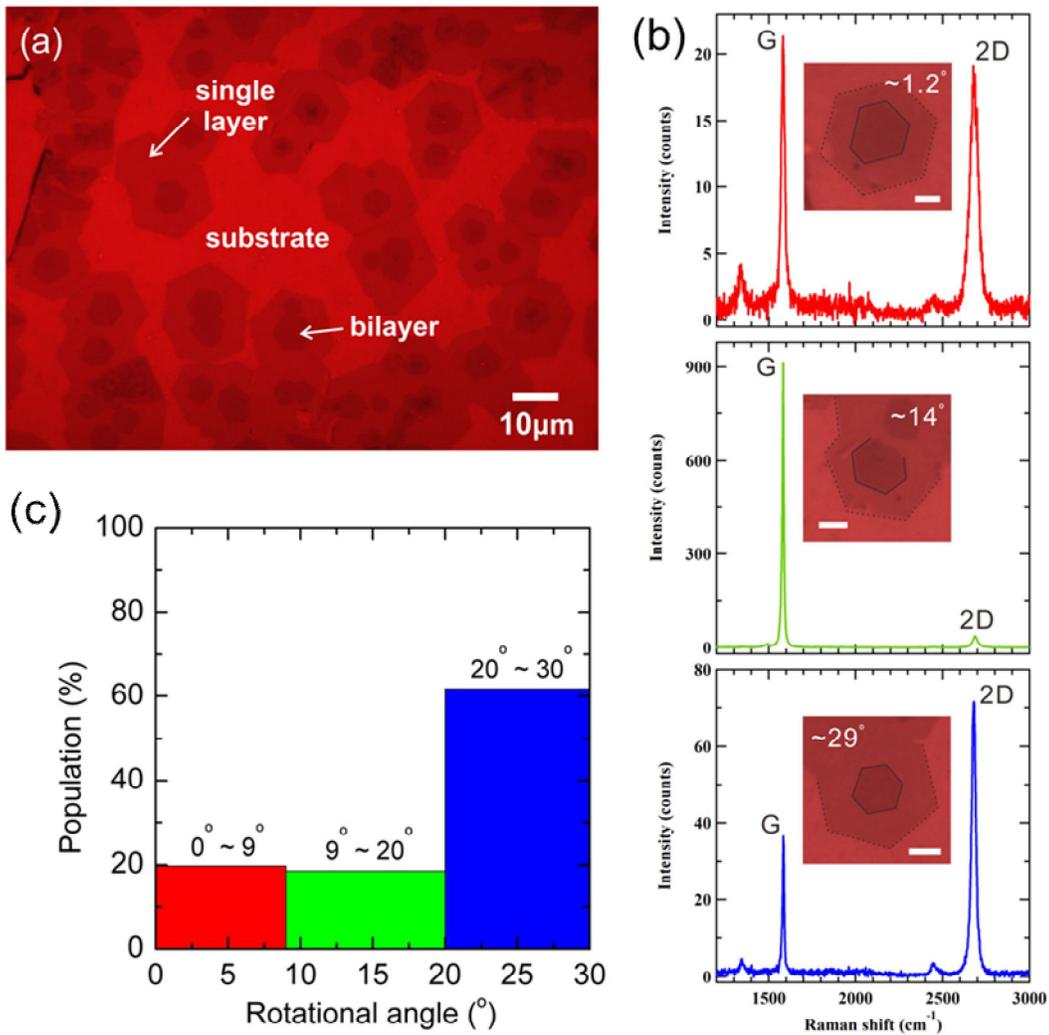

**Figure 1.** An example of twisted bilayer graphene (tBLG) domains grown by APCVD. (a) Optical image of tBLG domains transferred onto Si substrate with ~300 nm thermal oxide. The domains mostly have a hexagonal shape with edges parallel to zigzag directions.[17] (b) Representative Raman spectra and optical images of BLG domains with low, intermediate (near the critical angle where G peak intensity shows a resonance enhancement), and high twisting angles. The larger hexagonal first layer domains are highlighted by dashed lines, and the smaller hexagonal second layer domains are delineated by black solid lines for clarity. All Raman measurements were conducted at room temperature using a 532 nm laser excitation. The scale

bars in optical images of tBLG domains with twisting angles of ~1.2$^o$, ~14$^o$ and ~29$^o$ are 2, 3 and 5 μm, respectively. (c) Histogram of twisting angles of BLG domains in our CVD graphene sample determined by G and 2D Raman features. The histogram is based on a total of 81 BLG domains.

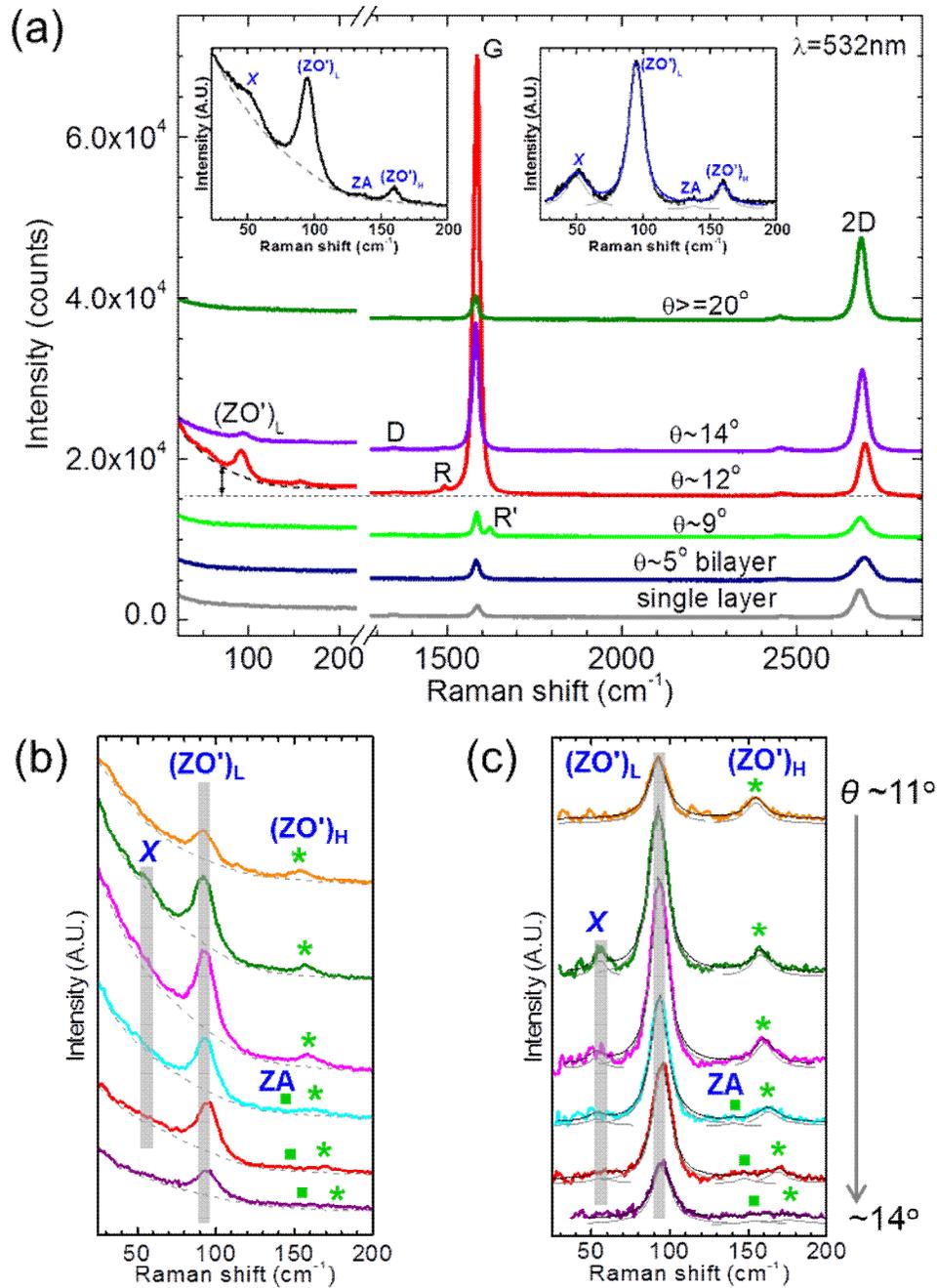

**Figure 2.** (a) Raman spectra from tBLG domains with different twisting angle $\theta$. A spectrum from SLG is included for comparison. The vertical scale is the same before and after the break on the horizontal axis. We define the low-energy background intensity to be the height intensity of the envelope (at 70 cm$^{-1}$) on which the low-energy Raman peaks superimpose (shown by the black vertical arrow for the spectrum with $\theta$~12°). The left inset displays a zoomed-in low-

energy spectrum. The right inset displays the same spectrum after subtraction of the background envelope highlighted by the dashed line shown in the left inset. The spectrum is decomposed into up to four Lorentzian peaks. (b) and (c) Original and background-subtracted low-energy Raman spectra from six different bilayer domains with twisting angles in the vicinity of $\theta_c$. Full spectra including the R, G and 2D bands are shown in the Supporting Information. Based on the R peak position, we determine that the twisting angle $\theta$ varies from ~11° to ~14°. In panel (c), the gray vertical bars highlight $X$ and $(ZO')_L$ modes. The squares and asterisks highlight ZA and $(ZO')_H$ modes, respectively. All spectra are excited by a 532 nm laser excitation.

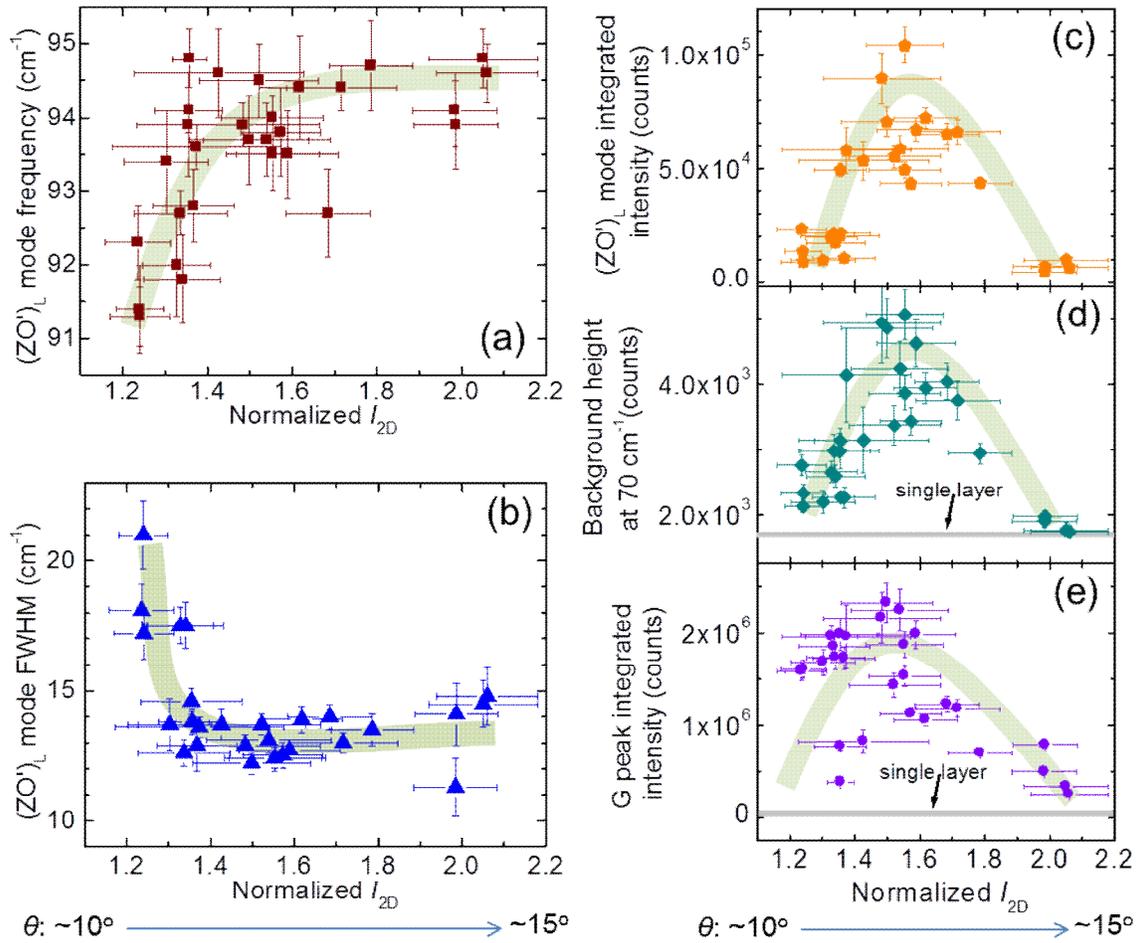

**Figure 3**. (a), (b) and (c) Frequency, full-width-at-half-maximum (FWHM), and integrated intensity of the $(ZO')_L$ mode as a function of normalized $I_{2D}$, respectively. The normalized $I_{2D}$ is defined as the ratio of the integrated 2D intensity of each bilayer domain to that of a single layer. The data are measured from a series of tBLG domains with twisting angle $\theta$ ranging from ~10° to ~15°, which corresponds to normalized $I_{2D}$ ranging from 1.1−2.2. (d) and (e) Background intensity at 70 cm$^{-1}$ (see Fig. 2(a) and its caption) and integrated intensity of the G peak as a function of the normalized $I_{2D}$, respectively. The horizontal lines in these two panels show the respective values of a single layer. The thick curves in each panel are guides to the eye. All results are obtained using a 532 nm laser excitation.

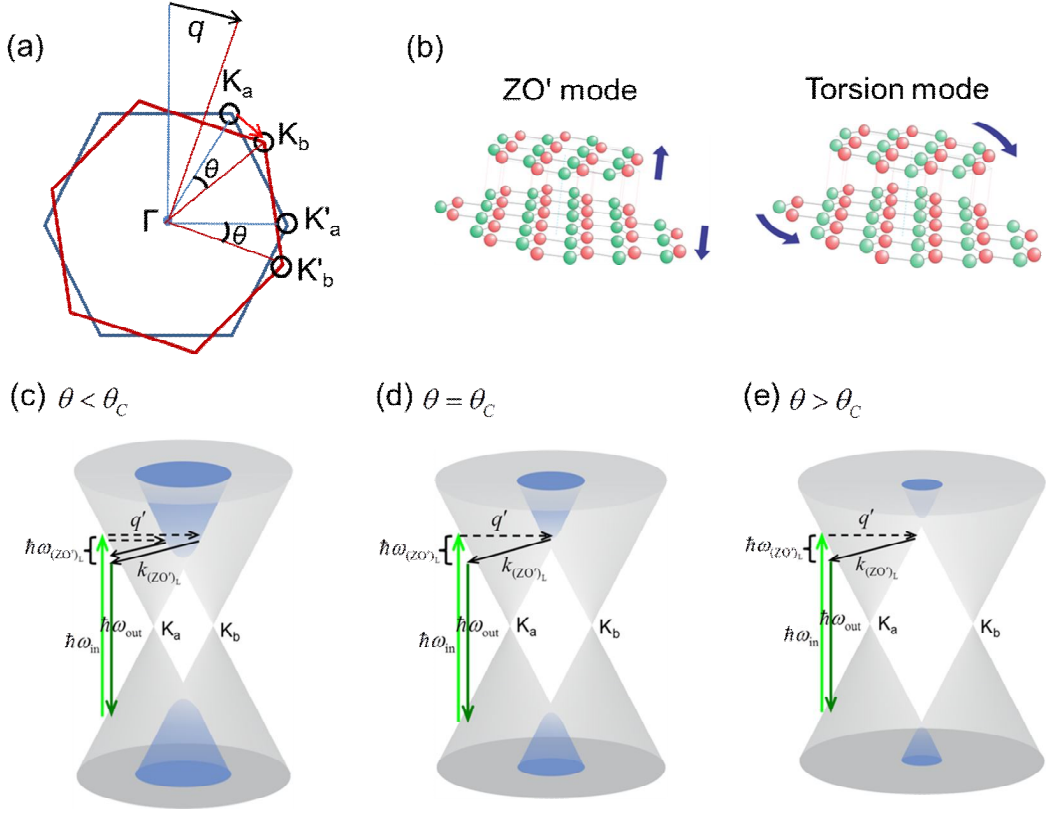

**Figure 4**. (a) The first Brillouin zone in the electronic band structure of tBLG with twisting angle $\theta$. $K_a$ and $K'_a$ are two adjacent Dirac points of the first graphene layer. $K_b$ and $K'_b$ are the two adjacent Dirac points of the second layer. $q$ is the wavevector of the tBLG superlattice (Moire pattern). (b) Schematic drawings of motions of atoms in the layer breathing (ZO') and torsion ($X$) modes. (c), (d), and (e) Schematic drawings of Raman processes of $(ZO')_L$ phonon when $\theta$ is less, equal to, or greater than the critical angle $\theta_c$, respectively. $\hbar\omega_{in}$ is the incident photon energy. $\hbar\omega_{out}$ is the scattered photon energy. $\hbar\omega_{(ZO')_L}$ is the phonon energy. The dashed arrows show the scattering of electrons by the tBLG crystal lattice. This scattering is elastic and is characterized by the wavevector $q'$. $k_{(ZO')_L}$ is the wavevector of the $(ZO')_L$ phonon. The portions of the two Dirac cones (at Dirac points $K_a$ and $K_b$) that overlap are shown in blue.

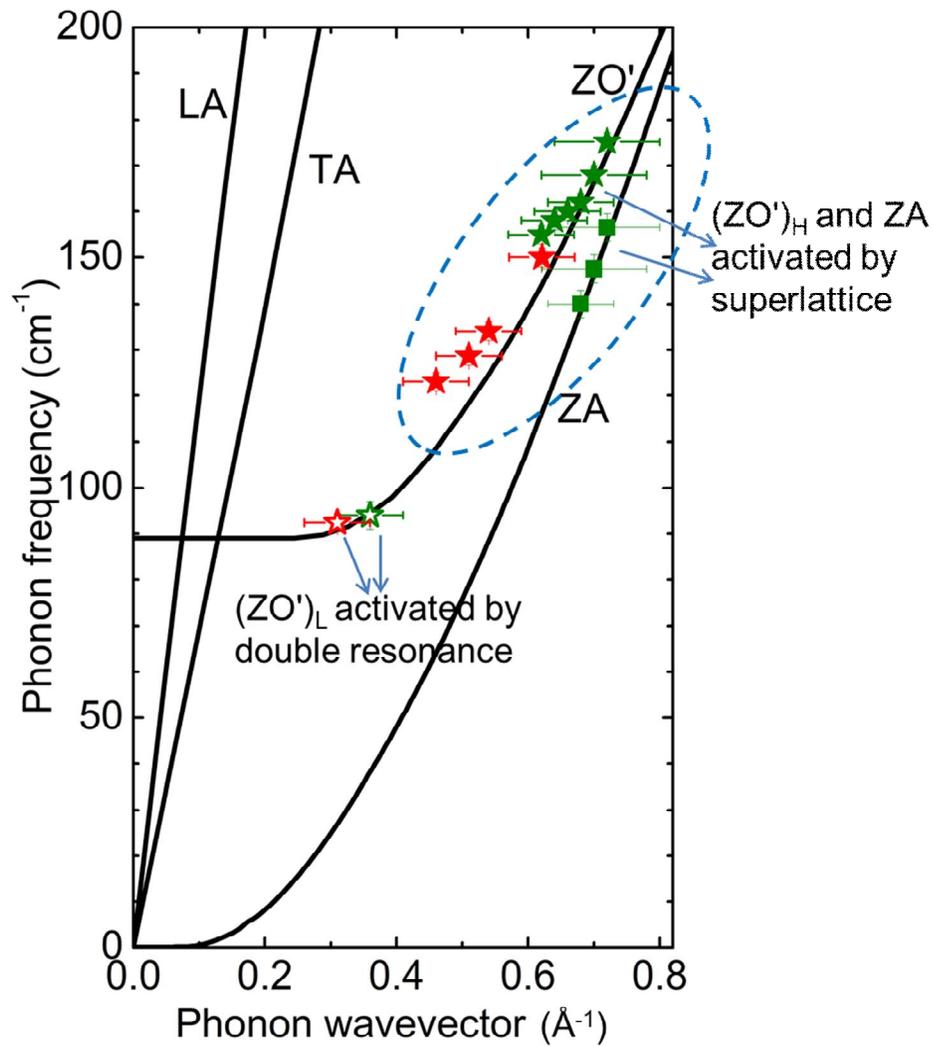

**Figure 5.** Low-frequency phonon dispersion. Different phonon branches are labeled. While the dispersion curves shown were calculated for SLG and Bernal-stacked bilayer graphene,[13] they are expected to be similar for tBLG for the relevant modes studied here. All points are determined by Raman measurements in this work. Points determined by a 532 nm laser excitation are shown in green, and those determined from a 633 nm laser excitation are shown in red.

# Supporting Information for:

# Observation of Low Energy Raman Modes in Twisted Bilayer Graphene

*Rui He,*[†,⊥,*] *Ting-Fung Chung,*[‡,⊤,⊥] *Conor Delaney,*[†] *Courtney Keiser,*[†] *Luis A. Jauregui,*[⊤,∥] *Paul M. Shand,*[†] *C. C. Chancey,*[†] *Yanan Wang,*[§] *Jiming Bao,*[§] *Yong P. Chen*[‡,⊤,∥,*]

[†] Department of Physics, University of Northern Iowa, Cedar Falls, IA 50614, USA

[‡] Department of Physics, Purdue University, West Lafayette, IN 47907, USA

[⊤] Birck Nanotechnology Center, Purdue University, West Lafayette, IN 47907, USA

[∥] School of Electrical and Computer Engineering, Purdue University, West Lafayette, IN 47907, USA

[§] Department of Electrical and Computer Engineering, University of Houston, Houston, TX 77204, USA

[⊥] These authors equally contributed to this work.

[*] Address correspondence to rui.he@uni.edu;　yongchen@purdue.edu

### 1. Experimental procedures

Graphene layers were grown by atmospheric pressure chemical vapor deposition (APCVD) on copper foil substrate (Alfa Aesar).[1-3] Growths of graphene domains were conducted in a tube furnace at 1050−1065 °C for 15 min. Copper foil surfaces were pre-annealed at 1050 °C for 3 hours before growth. Two different growth parameters were applied: (i) 100 sccm $CH_4$, 170 sccm Ar, and 30 sccm $H_2$. (ii) 40 sccm $CH_4$, 437 sccm Ar, and 23 sccm $H_2$. The morphology of the graphene bilayer domains looks slightly different under the two growth conditions. The first recipe produces more hexagonal-shape bilayers. However, no noticeable difference in the Raman spectra was observed in graphene bilayer domains grown by these two recipes.

Graphene layers were transferred to a highly p-doped Si substrate (with 300 nm $SiO_2$) using the following procedures: PMMA (poly(methylmethacrylate), A4) was spin-coated on the copper/graphene surface. Then we used oxygen plasma etching to remove the graphene layers on the back side of the Cu foil. The PMMA-coated copper/graphene was then placed in a dilute aqueous $FeCl_3$ solution to etch the copper off. The PMMA/graphene membrane floating on the etchant solution was washed with copious amount of water. After the cleaning process, the membrane was scooped by a $Si/SiO_2$ wafer and dried in air for a couple of hours. In the end, the sacrificial PMMA layer was dissolved using acetone followed by rinsing with IPA.

Raman spectra were taken at room temperature in ambient conditions using a Horiba Xplora (equipped with 532 nm laser light) and a Horiba Labram HR Raman Microscope system (equipped with 532 and 633 nm lasers). Both spectrometers have thermo-electric cooled charge-coupled device (CCD) detectors. For the Horiba Labram HR Raman Microscope system, a 532-nm Rayleigh filter enables access to ultra-low frequencies down to 20 $cm^{-1}$. For 633 nm, the lowest frequency that can be accessed by our spectrometers is about 70 $cm^{-1}$. Incident laser power was kept below 5 mW and the acquisition time was 10-15 seconds. A 100× objective lens was used. Spectral resolution was about 1 $cm^{-1}$.

## 2. Atomic Force Microscopy (AFM) characterization of BLG domains

The number of layers is confirmed by measuring the thickness of bilayer domains by AFM. Measurement was performed in the tapping mode at room temperature in air using a NT-MDT Ntegra Prima system. A representative AFM image is shown in Fig. S1(a). Step height profiles of AFM images (see Figs. S1(b) and (c)) show that the thicknesses of the first (bottom) and the second (top) layers are ~1.5 nm and ~0.5 nm, respectively. The deviations from the expected single-layer graphene (SLG) thickness and interlayer spacing (~0.35 nm per layer[4]) could be due to adsorbed molecules on the surface of graphene and between the graphene and the $SiO_2$ substrate, transfer-induced wrinkles, tip-substrate interaction,[5] and different stacking order in twisted BLG. In Fig. S1(a) we see that there are many impurities (seen as bright spots in the image) on the surface of the sample. These impurities are probably PMMA residues due to PMMA-assisted transfer. These residues usually turn graphene into p-type doping,[6] similar to the effect of water vapor and oxygen.[7] The similarities in the G and 2D Raman features between our samples (see Figs. S2, S3, and S10) and those obtained in suspended tBLG[8] suggest that doping due to the impurities and substrate does not have substantial effect on our Raman spectra. In addition, the residues and impurities are on the surface of the samples and not in between the two graphene layers which are in full contact and grown naturally by CVD.[9] Therefore, these impurities are not expected to have a strong impact on the low-energy vibrational modes that arise from the relative or in-phase motion of the two graphene layers.

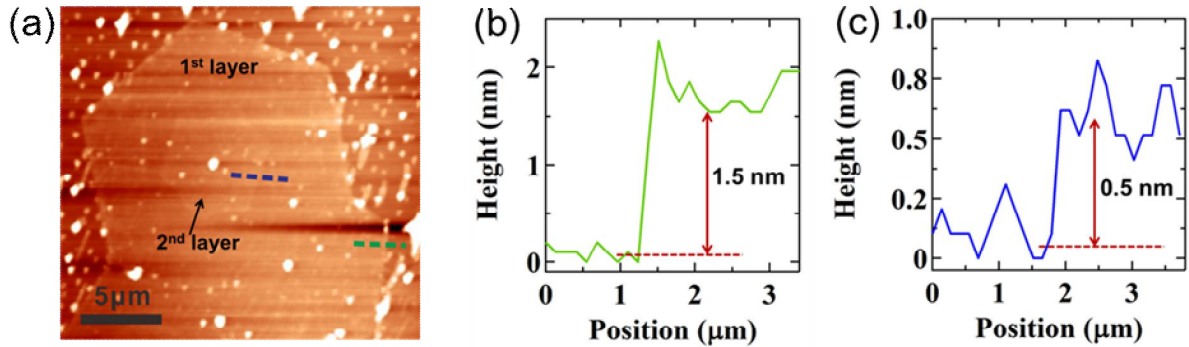

Figure S1. (a) AFM image of a bilayer domain. The second (top) layer shows a weaker color contrast compared to the first (bottom) layer. The bright spots are from impurities or residues introduced during transfer. (b) Height profile measured along the dotted green line shown in panel (a). It shows the height of the first graphene layer with respect to the substrate surface. (c) Height profile measured along the dotted blue line shown in panel (a). It shows the height of the second graphene layer with respect to the first layer.

## 3. Dependence of G and 2D peaks on twisting angles of bilayer domains

Figures S2 and S3 show the features of G and 2D Raman bands (excited by 532 nm laser light) from BLG domains synthesized by the first and second recipes, respectively. Samples grown by the two recipes show similar Raman characteristics. The rotational angles were estimated by comparing the G and 2D Raman features with those published in Refs. [9] and [8]. Similar to the prior work, the maximum of integrated intensity of the G Raman band is found to be at a rotational angle of ~12.5°, and the normalized intensity of the 2D band (normalized to the single layer value) grows monotonically with rotational angle in the range of 10°–16°. Figure S4 shows the G and 2D Raman features as functions of rotational angle determined by the method of edge misorientation (measuring angles between hexagonal edges of the two layers) for samples grown by recipe (i). We can see that the twisting angle estimations by both methods are consistent (difference is less than 2°).

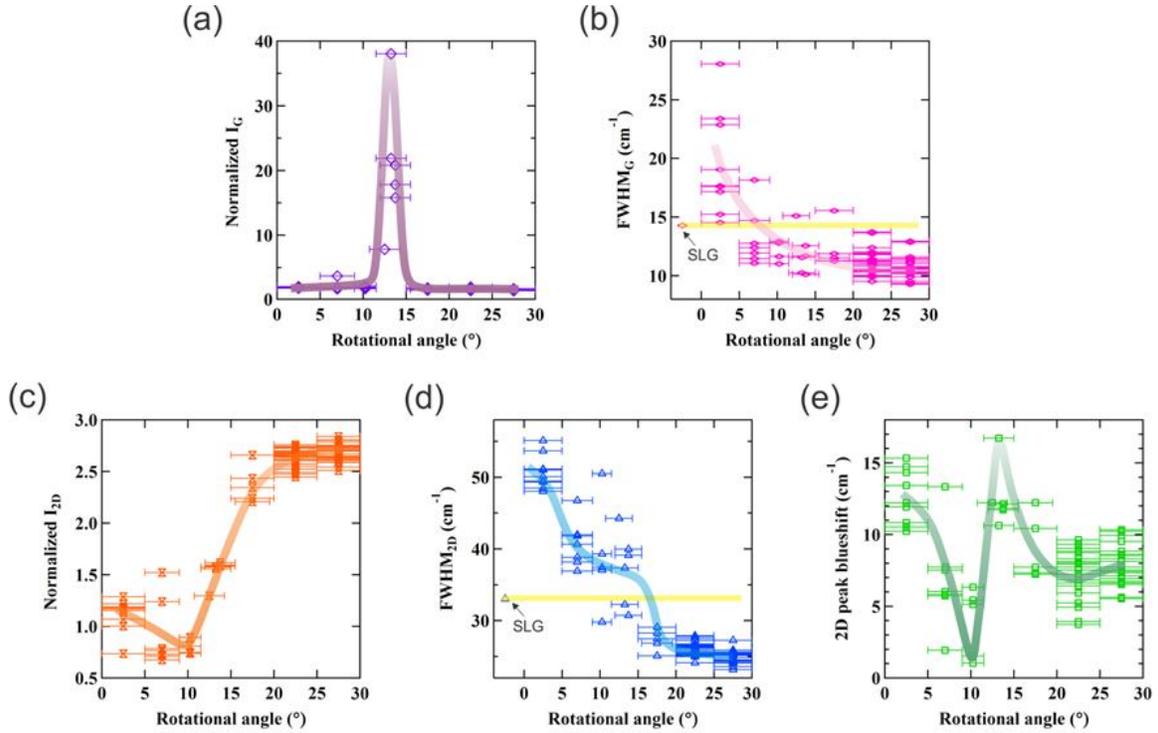

Figure S2. Rotational-angle dependence of G and 2D Raman features from bilayer graphene domains grown by recipe (i). A total of 81 bilayer domains were characterized. (a) Integrated intensity of the G band (normalized to the value of SLG). (b) FWHM of the G band. (c) Integrated intensity of the 2D band (normalized to the SLG value). (d) FWHM of the 2D band. (e) 2D band position (blueshift with respective to the value of SLG). Rotational-angles were categorized into 7 groups (0-5°, 5-9°, 9-11.5°, 11.5-15°, 15-20°, 20-25°, 25-30°) and determined by the characteristics of Raman peaks (G, 2D, R and R').[8-10] The colored curves are guides to the eye. The horizontal lines show the experimental values for SLG. All Raman measurements were conducted at room temperature using a 532 nm laser excitation.

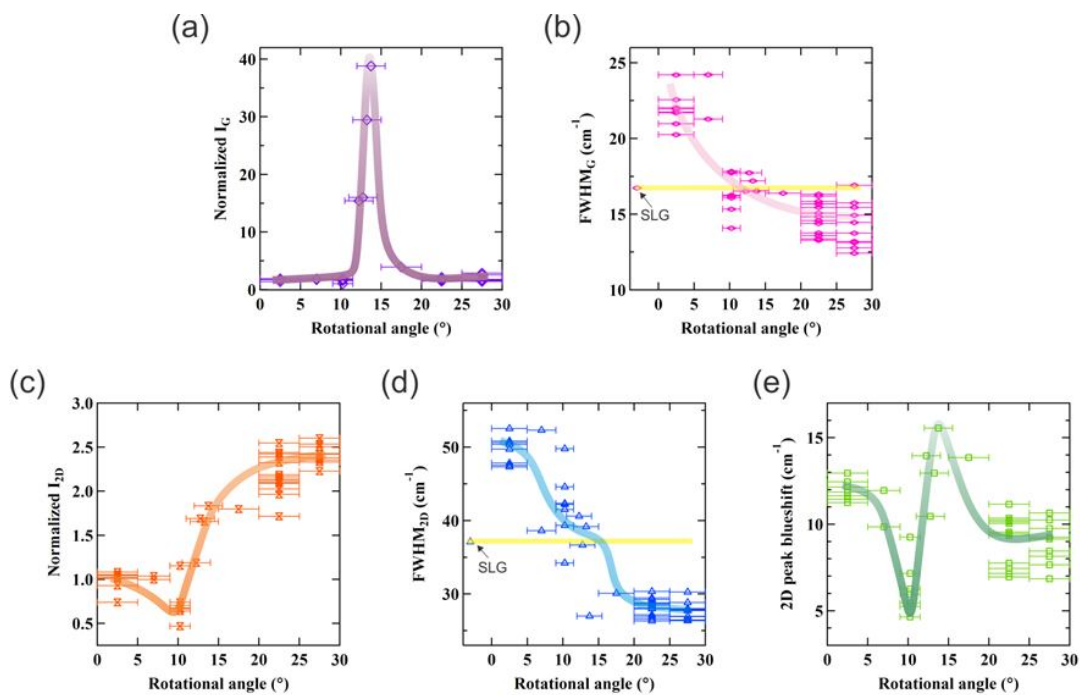

Figure S3. Same as Fig. S2 for bilayer domains grown by recipe (ii). A total of 47 domains were characterized.

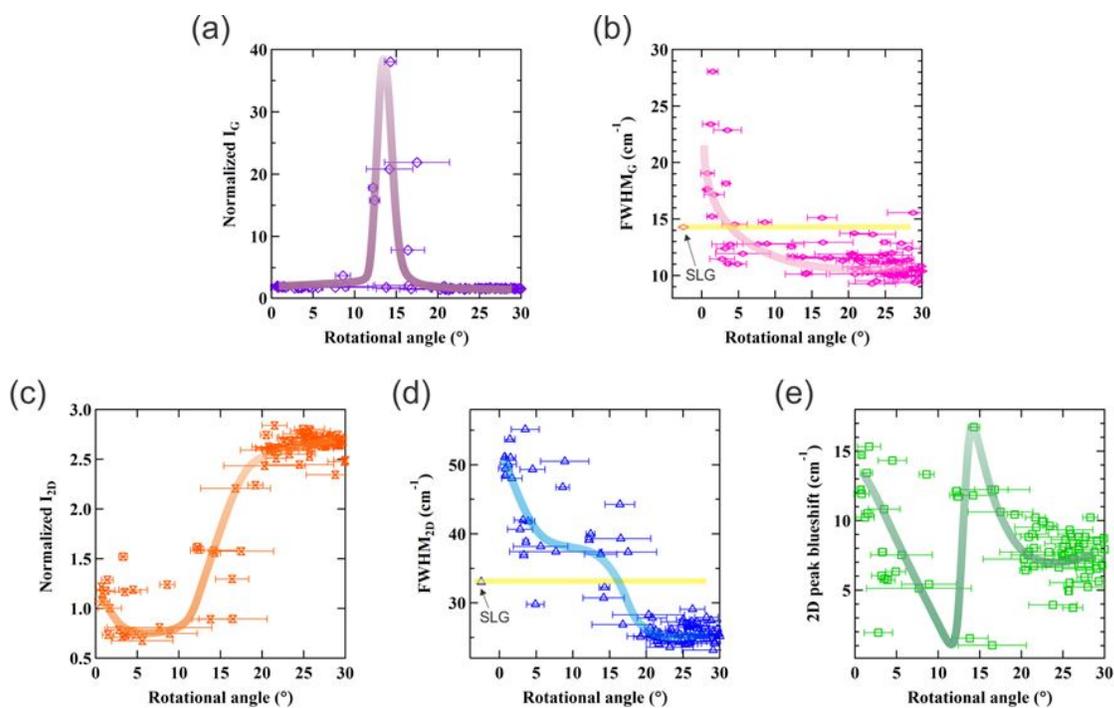

Figure S4. Same as Fig. S2 for the same 81 bilayer domains grown by recipe (i). However, the rotational angles in this figure were estimated by measuring the angles formed between neighboring edges of the first and the second layer domains (examples are shown in the insets of Fig. 1(b) in the paper).

Figure S5 summarizes the distribution of twisting angles of BLG domains (grown by recipe (i)) determined by the relative orientations between the edges of top and bottom graphene layers seen in optical images (examples are shown in the insets of Fig. 1(b) in the paper). It is consistent with the distribution of twisting angles determined by the Raman method (see Fig. 1(c) in the paper).

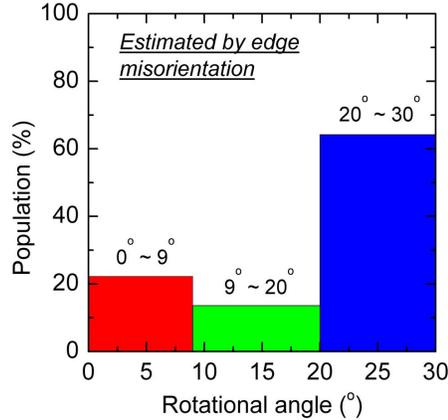

Figure S5. Histogram of BLG twisting angles determined by the relative orientations between the edges of top and bottom graphene layers seen in optical images (examples are shown in the insets of Fig. 1(b) in the paper). This histogram and the one shown in Fig. 1(c) are based on a total of 81 BLG domains (all grown by recipe (i)) whose twisting angles are characterized by both methods.

Figure S6 shows the distribution of twisting angles of BLG domains grown by recipe (ii) determined by Raman method. It is consistent with that of BLG domains grown by recipe (i) as shown in Fig. 1(c) in the paper.

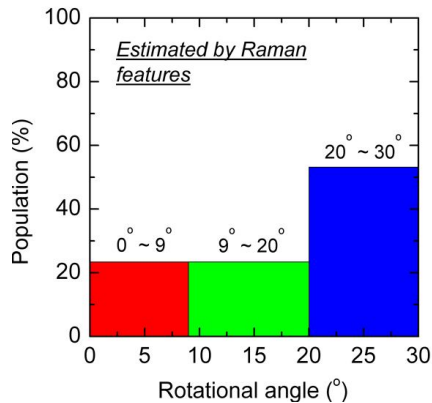

Figure S6. Another histogram of twisting angles determined by G and 2D Raman features for the BLG domains grown by recipe (ii). A total of 47 domains were investigated. All measurements were performed in ambient condition using a 532 nm laser excitation.

## 4. Additional analysis of Raman spectra from tBLG

Figure S7 shows full range Raman spectra of those shown in Figs. 2(b) and (c) in the paper. In particular, we have used the R peak position (following Ref. [11]) as another method to determine the twisting angles in tBLG.

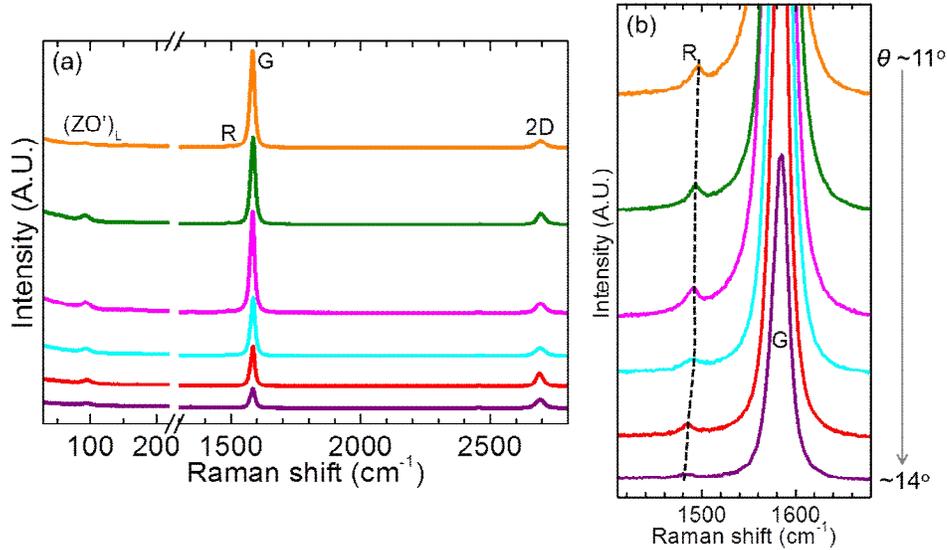

Figure S7. (a) Full range Raman spectra of those shown in Figs. 2(b) and (c) in the paper. (b) Zoomed-in spectra in the region of R peak. For the spectra from top to bottom: the R peak positions are at 1496, 1494, 1492, 1490, 1485, 1483 cm$^{-1}$, respectively, and $\theta$ varies from ~11° to ~14°. All spectra were measured by a 532 nm excitation laser.

Figure S8 shows the frequency differences of the G and R peaks from tBLG domains and compares them with the frequencies of the $(ZO')_L$ mode. In this figure we can see that $\omega_G - \omega_R$ is not correlated with $\omega_{(ZO')_L}$.

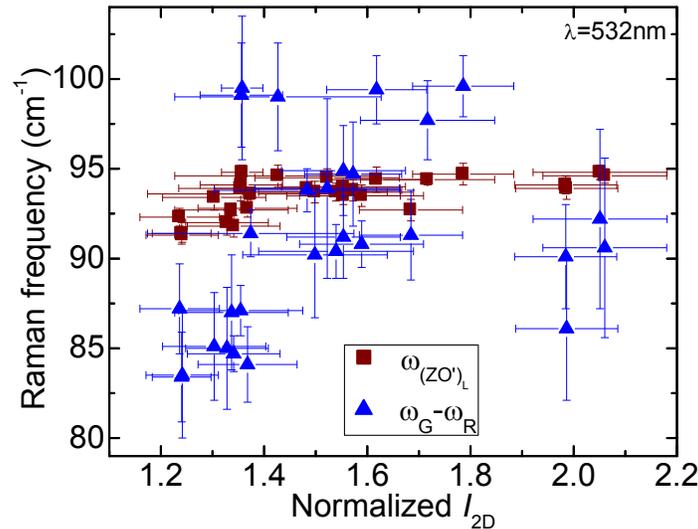

Figure S8. Comparison of $(ZO')_L$ frequencies with the frequency differences between G and R peaks from tBLG domains.

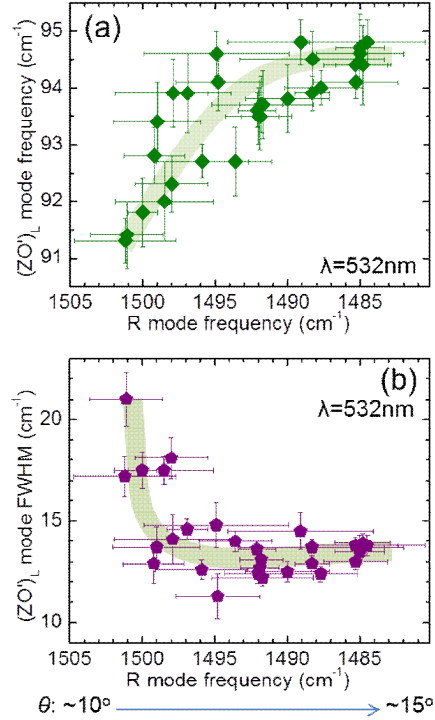

Figure S9. Frequency and FWHM of the $(ZO')_L$ mode as a function of the R mode frequency. The thick curves in each panel are guides to the eye.

Figure S10 shows the dependence of G peak frequency on the normalized 2D intensity from tBLG domains. The G peak frequencies agree well with those from suspended tBLG samples reported in Ref. [8], which suggests that doping from the substrates in our samples is not a major concern.

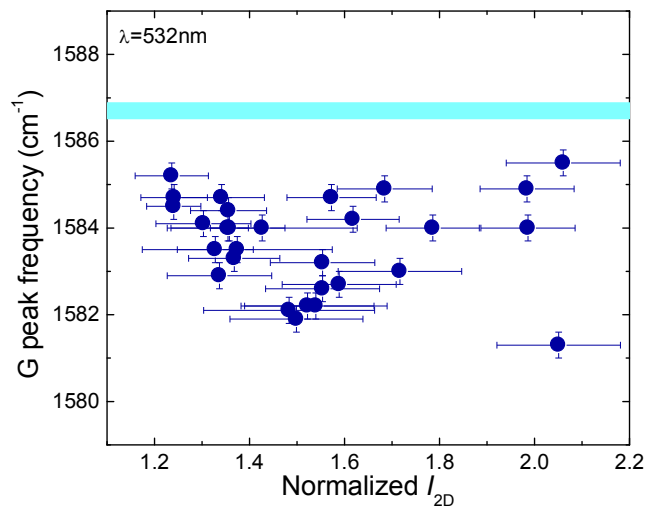

Figure S10. Dependence of G peak frequencies on the normalized $I_{2D}$ from tBLG domains. The horizontal line shows the G frequency of SLG.

We also used 633 nm laser light to probe the low energy modes from tBLG. Similar twisting-angle-dependent changes in the Raman spectra (see Fig. S11) as those seen under 532 nm laser excitation are observed. The critical twisting angle $\theta_c$ is ~10° for 633 nm excitation.[8] Because our spectrometer does not access Raman modes below 70 cm$^{-1}$ under 633 nm laser excitation, the spectra in Figs. S11 and S12 stop at this frequency at the low energy end. The $(ZO')_L$ and $(ZO')_H$ modes are indicated in Fig. S12(a). The background envelopes shown by the gray dashed lines in Fig. S12(a) are estimations since the cutting edge of 70 cm$^{-1}$ overlaps the low energy tail of the $(ZO')_L$ mode. Subtraction of background envelopes will have large uncertainty. Therefore, only the frequencies of the $(ZO')_L$ mode were extracted by taking the positions of this peak (as shown by the triangles in Fig. S13). Bandwidths and integrated intensities which involve background subtraction and lineshape fittings are not conducted since they will have large uncertainty. In Fig. S13 we see that the $(ZO')_L$ phonon frequency displays redshift slightly under 633 nm laser excitation in comparison to that excited by 532 nm laser light. This phonon softening confirms that the $(ZO')_L$ phonon has a nonzero wavevector and that the scattering process can be explained by the double resonance mechanism. The softening at longer laser excitation wavelength (lower excitation photon energy) is consistent with the $(ZO')_L$ phonon dispersion shown in Fig. 5 in the paper.

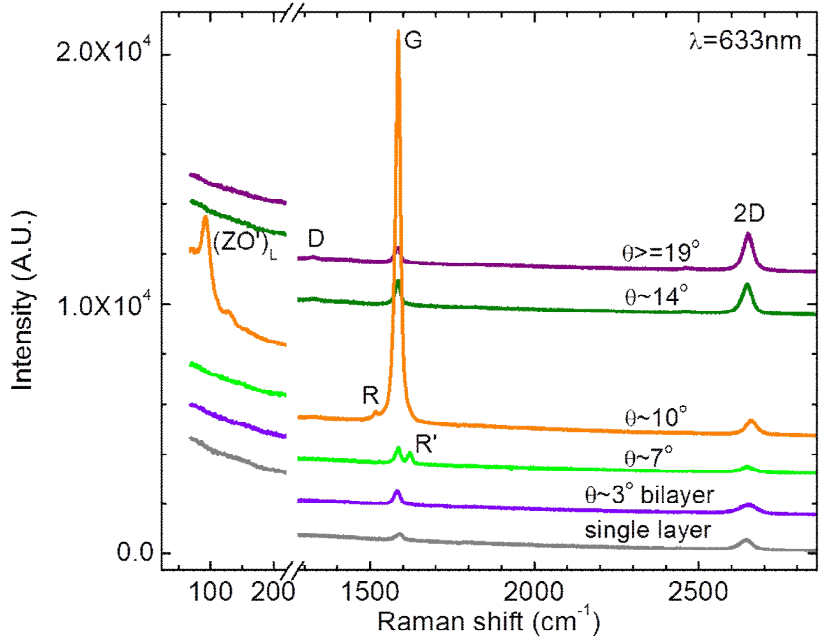

Figure S11. Raman spectra from graphene bilayer domains with different twisting angles $\theta$. A spectrum from SLG is included for comparison. The vertical scale is the same before and after the break on the horizontal axis. Because of the Rayleigh filter used for 633 nm laser excitation in our system (different from the one for 532 nm laser light), the lowest frequency we can access is ~70 cm$^{-1}$, and the spectra show stronger stray light intensity which caused a steeper rising background toward the low-frequency end of the spectra in comparison to those obtained under 532 nm laser excitation (see Fig. 2(a) in the paper). As a result, the spectra excited by 633 nm laser light show larger vertical gaps at the place where the horizontal axis breaks.

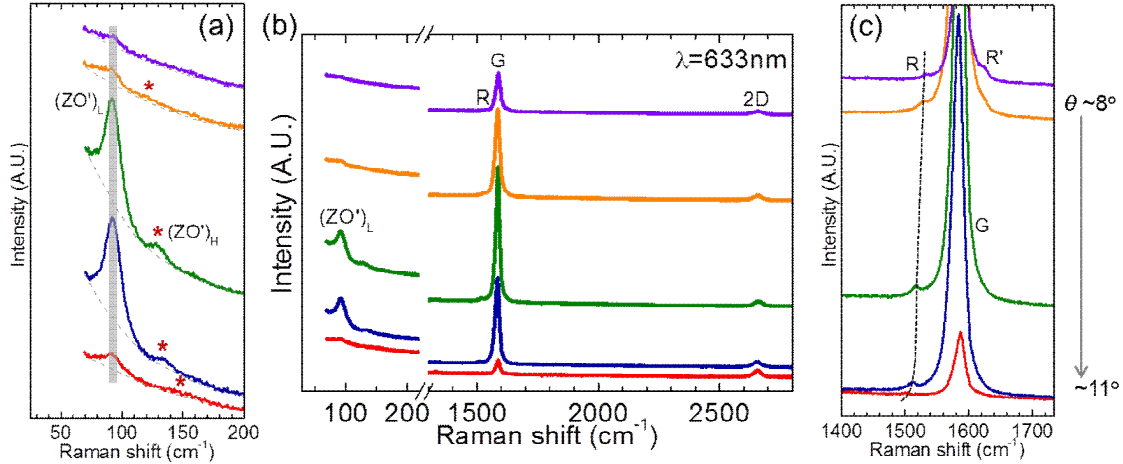

Figure S12. (a) Raman spectra of low-lying modes from 5 different bilayer domains that have twisting angle $\theta$ in the vicinity of $\theta_c \approx 10°$. The spectra are measured with 633 nm laser light. The dashed gray lines show the background envelopes. Positions of the $(ZO')_L$ and $(ZO')_H$ modes are highlighted by the vertical bar and asterisks, respectively. (b) Full range Raman spectra of those shown in panel (a). (c) Zoomed-in spectra in the region of R peak. For the spectra from top to bottom: the R peak positions are at 1535, 1530, 1517, 1513, and 1501 cm$^{-1}$, respectively, and $\theta$ varies from ~8° to ~11°.

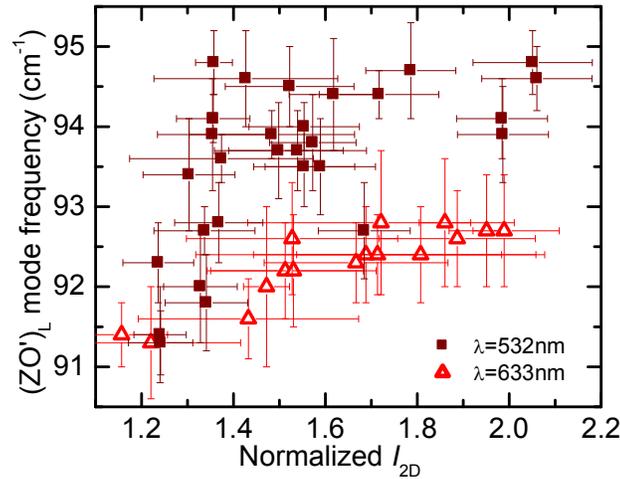

Figure S13. Frequency of the $(ZO')_L$ mode as a function of normalized 2D intensity under 532 and 633 nm laser excitations. The range of 1.1−2.2 for normalized 2D intensity corresponds to a range of twisting angle $\theta$ from ~10° to ~15° for 532 nm laser light, and from ~9° to 14° for 633 nm laser light.

## 5. Raman maps of the (ZO')$_L$ mode from a tBLG domain

Figure S14 shows Raman maps of the (ZO')$_L$ mode and the background height from a tBLG domain that has a twisting angle $\theta \sim 12.5^o$. Under the 532 nm laser light, the (ZO')$_L$ mode is well-defined, and a typical Raman spectrum from this domain is shown in Fig. S7 (the magenta line or third spectrum from the top). Raman maps of the frequency, FWHM, and integrated intensity of the (ZO')$_L$ mode and map of the background intensity are shown in Fig. S14(b)-(e). Because of shorter acquisition time, the integrated intensity of this (ZO')$_L$ mode at each point in the Raman map is lower than those shown in Figs. 2 and 3 in the paper. Under the 633 nm laser excitation, the (ZO')$_L$ mode from the same domain is hardly observed, as evidenced in Fig. S14(f).

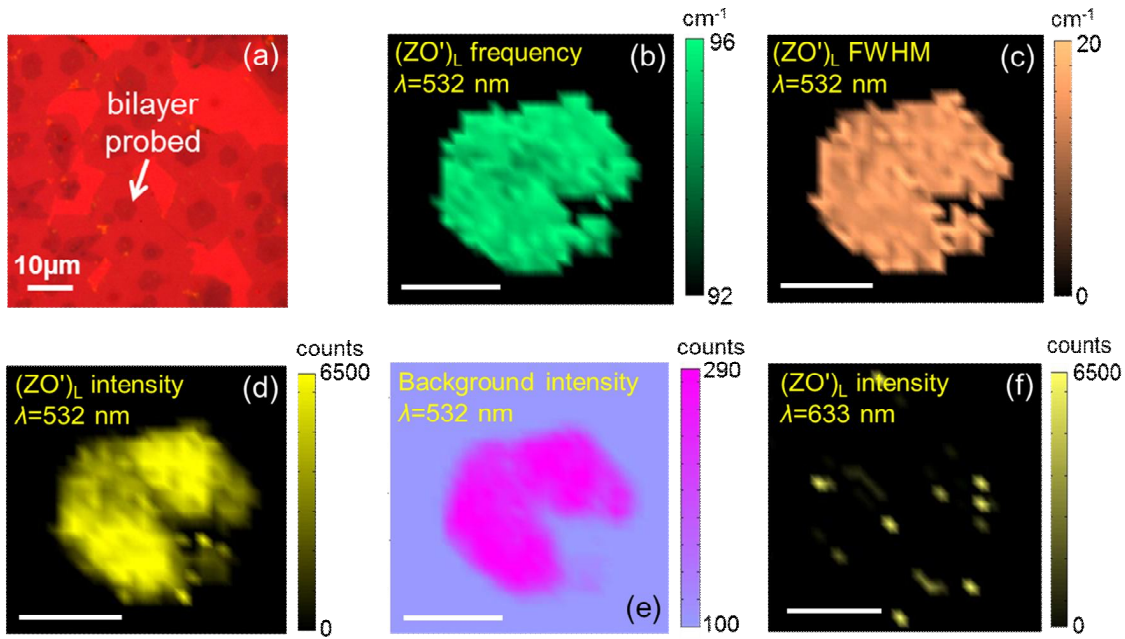

Figure S14. (a) Optical image of tBLG domains. The arrow indicates the one that is probed by Raman mapping. (b)-(e) Raman maps of the frequency, FWHM, intensity of the (ZO')$_L$ mode, and the background height at 70 cm$^{-1}$. The excitation laser wavelength is 532 nm. (f) Raman map of the intensity of the (ZO')$_L$ mode from the same domain by using a 633 nm laser excitation. In panels (b)-(f), the scale bars are all 3 μm. Little signal is detected from the lower right corner of the island, probably due to damage in that area.